\documentclass[prd,nofootinbib,12pt]{revtex4}

\usepackage{lineno,hyperref}
\usepackage{amsfonts,amssymb,amsmath}
\usepackage{epsfig}
\usepackage{mathrsfs}
\usepackage{amssymb}
\usepackage{graphicx}
\usepackage{amsmath}
\usepackage{graphicx}
\usepackage{amsmath}
\usepackage{xcolor}
\usepackage{color}
\usepackage{appendix}

\begin{document}

\title{Inflationary power spectrum from the Lanczos algorithm}

\author{Ke-Hong Zhai$^{1}$}
\author{Lei-Hua Liu$^{1}$}
\email{liuleihua8899@hotmail.com}
\affiliation{$^1$Department of Physics, College of Physics, Mechanical and Electrical Engineering, Jishou University, Jishou 416000, China}
\author{Hai-Qing Zhang$^{2}$}\email{hqzhang@buaa.edu.cn}
\affiliation{Center for Gravitational Physics, Department of Space Science, Beihang University, Beijing
	100191, China}
\affiliation{Peng Huanwu Collaborative Center for Research and Education, Beihang University, Beijing 100191, China}

\begin{abstract}

The generalized Lanczos algorithm can provide a universal method for constructing the wave function under the group structure of Hamiltonian. Based on this fact, we obtain an open two-mode squeezed state as the quantum origin for the curvature perturbation. In light of this wave function in the open system, we successfully develop a new method to calculate its corresponding power spectrum by using the Bogoliubov transformation. Unlike traditional approaches, we explicitly retain the Bogoliubov coefficients in terms of the squeezing amplitude \( r_k \) and the squeezing rotation angle \( \phi_k \). As a result, the power spectrum of the open two-mode squeezed state will match that of the Bunch-Davies vacuum numerically. Furthermore, the derivation of the open two-mode squeezed state relies on the second kind Meixner polynomial (equivalent to the generalized Lanczos algorithm) and the symmetry of the Hamiltonian. Therefore, our research may offer a new insight into the calculation of the correlation functions through a group-theoretic perspective.

\end{abstract}

\maketitle


\section{introduction}
\label{introduction}

The inflationary paradigm \cite{Guth:1980zm,Starobinsky:1980te,Linde:1981mu} addresses several fundamental issues in standard Big Bang cosmology, including the flatness problem, the horizon problem, and the magnetic monopole problem. Quantum fluctuations in the vacuum state during quasi-de Sitter expansion predict the power spectrum of primordial curvature perturbations \cite{Mukhanov:1990me,Martin:2024qnn}, which is nearly scale-invariant as constrained by observations of the cosmic microwave background (CMB) \cite{Planck:2018jri}. The advent of high-precision CMB measurements has initiated an era of precision cosmology. Consequently, current and forthcoming observations are capable of probing primordial curvature perturbations beyond the predictions of standard inflationary models. \cite{CORE:2016ymi,CMB-S4:2020lpa,LiteBIRD:2022cnt,Euclid:2023shr,Mergulhao:2023ukp,Antony:2024vrx}. 

The quantum origin of curvature perturbations has attracted significant attention \cite{Balitsky:2014epa, Dalianis:2024}. This motivates the application of alternative methods from quantum information theory to evaluate curvature perturbations. One promising direction involves utilizing the concept of complexity to study these perturbations. Several types of complexity exist in quantum information, including computational complexity \cite{Papadimitriou:2003}, Krylov complexity \cite{Parker:2018yvk}, among others. Computational complexity in the context of inflation has been investigated in \cite{Choudhury:2020hil, Bhargava:2020fhl, Bhattacharyya:2020rpy, Adhikari:2021ked, Li:2023ekd, Liu:2021nzx, Li:2021kfq}. Among these, Ref. \cite{Adhikari:2022oxr} studied Krylov complexity in cosmology with non-trivial sound speed. However, these works primarily address closed quantum systems. Since our universe is fundamentally an open system, it is essential to adopt an open quantum systems approach to study Krylov complexity during inflation \cite{Li:2024kfm, Zhai:2024odw}. In \cite{Li:2024kfm, Zhai:2024odw}, the authors successfully employed the two-mode squeezed state formalism to analyze Krylov complexity in inflation. As emphasized in \cite{Grishchuk:1990bj}, the amplification of quantum fluctuations to classical scales during inflation is effectively a quantum squeezing process. Therefore, the two-mode squeezed state provides a natural framework for describing quantum fluctuations in this context. This approach has also been extensively applied to study the pre-inflationary phase \cite{Kowalczyk:2024ech, Vilenkin:1982wt, Starobinsky:2001kn, Gessey-Jones:2021yky, Handley:2016ods, Contaldi:2003zv}.

In this work, we employ the Bogoliubov transformation to relate the two-mode squeezed state to the Bunch-Davies (BD) vacuum. The Bogoliubov coefficients, $\alpha_k$ and $\beta_k$, are determined according to \cite{Bunch:1978yq}. The wave function used here is not the conventional pure two-mode squeezed state, but rather the open two-mode squeezed state (OTMSS), constructed via the second kind of Meixner polynomial, which is equivalent to the generalized Lanczos algorithm along with the symmetries of the Hamiltonian \cite{Zhai:2024odw}. Within this framework, we compute the power spectrum of the OTMSS using the generalized Lanczos coefficients. Our numerical results show agreement with those obtained in the BD vacuum.

This work is organized as follows. Sec. \ref{Some foundation of inflationary perturbation} will give some foundations of curvature perturbation under the framework of single-field inflation. Sec. \ref{Lanczos algorithm in inflation} describes the generalized Lanczos algorithm in inflation. In Sec. \ref{wave function}, we will review the so-called OTMSS. Sec. \ref{Bogoliubov transformation} will show the explicit formula of $\alpha_k$ and $\beta_k$. Then, Sec. \ref{Evolution of phik and rk} will obtain the numerics of $\phi_k$ and $r_k$. Based on this, Sec. \ref{Power spectrum of open tw-mode squeezed state} will derive the power spectrum of OTMSS. Finally, the conclusion and discussion will be given in Sec. \ref{conclusion and outlook}.


\section{Some foundation of inflationary perturbation}
\label{Some foundation of inflationary perturbation}
In this work, we will follow Ref. \cite{Baumann:2009ds} to give some foundation for the cosmological perturbation. First, it is the spatially conformal flat Friedmann-Lemaitre-Robertson-Walker (FLRW) metric,
\begin{equation}
ds^2=a^2(\eta)[-(1+\psi(\eta,\vec{x}))d\eta^2+(1-\psi(\eta,\vec{x}))d\vec{x}^2],
\end{equation}
where $\psi(\eta,\vec{x})$ denotes the metric perturbation and $a(\eta)$ is the scale factor. We also define the perturbation of the inflaton field as $\phi(x_{\mu})=\phi_0(\eta)+\delta\phi(x_\mu)$, where $\phi_0(\eta)$ is the homogeneous background inflaton and $\delta\phi(x_\mu)$ represents the inflaton perturbation. Using these definitions, the generic quadratic action can be derived as follows:
\begin{equation}
S=\frac{1}{2}\int dtd^3xa^3\frac{\dot{\phi}}{H^2}\bigl [\dot{\mathcal{R}}-\frac{1}{a}(\partial_i\mathcal{R})^2\bigr ],
\label{eq S}
\end{equation}
where the Hubble parameter is defined as $H=\dot{a}/a$ (denoting the derivation with the physical time), the conserved curvature perturbations is $\mathcal{R}=\psi+\frac{H}{\phi_{0}}\delta\phi$ at horizon crossing $(k=aH)$ and the  slow-roll parameter is $\epsilon=-\dot{H}/H^2$, which leads to  $z=\sqrt{2\epsilon}a$. Furthermore, the action \eqref{eq S} can be rewritten in terms of the Mukhanov-Sasaki variable $v=z\mathcal{R}$,
\begin{equation}
S=\frac{1}{2}\int d\eta d^3x[v'^2-(\partial_iv)^2+(\frac{z'}{z})^2v^2-2v'v\frac{z'}{z}].
\label{quadratic action in terms of v}
\end{equation}
Applying the Euler-Lagrange equation, we obtain the equation of motion in the momentum space,
\begin{equation}
v_k''+(k^2-\frac{z''}{z})v_k=0.
\label{eom of v}
\end{equation}
One possible solution of Eq. \eqref{eom of v} is the well-known BD vacuum,
\begin{equation}
v_{BD}(\eta,\vec{k})=\frac{e^{-ik\eta}}{\sqrt{2k}}(1-\frac{i}{k\eta}).
\label{BD vacuum}
\end{equation}
However, the definition of a privileged quantum vacuum state in curved spacetime is fundamentally ambiguous, as emphasized in \cite{Unruh:1976db}. In this setting, the BD vacuum arises as a specific state that approximates the Minkowski vacuum at short distances \cite{Agullo:2022ttg}. A complete characterization of the vacuum requires the framework of the Poincaré patch of de Sitter spacetime (PdS) together with Hadamard regularity conditions. In accordance with standard cosmological practice, we adopt the BD vacuum under the assumptions of PdS invariance and Hadamard regularity, following the foundational treatments in \cite{Chernikov:1968zm,Tagirov:1972vv,Bunch:1978yq}.

From an operational perspective, establishing the BD vacuum requires two key elements:
$(a)$. The initial conditions for Equation \eqref{eom of v} in the sub-horizon regime preceding inflation. 
$(b)$. The harmonic approximation of \eqref{eom of v} as a quantum harmonic oscillator.

The initial condition of Eq. \eqref{eom of v} is that it can be approximated as a quantum harmonic oscillator in sub-horizon regime before inflation.
These conditions demonstrate that the BD vacuum represents a highly specialized choice for inflationary quantum perturbations. We propose that a more fundamental vacuum state might exist, derivable through Hamiltonian symmetry principles and the Lanczos algorithm. This approach could reveal deeper structure in the quantum initial conditions of inflation.

\section{Lanczos algorithm in inflation}
\label{Lanczos algorithm in inflation}
Recent studies \cite{Li:2024kfm,Li:2024iji,Li:2024ljz,Zhai:2024odw,Shandera:2017qkg} have demonstrated that the early universe behaved as a dissipative system, owing to both the violation of energy conservation \cite{Cheung:2007st} and the particle production during preheating \cite{Kofman:1997yn,Kofman:1994rk}. Motivated by these findings, we employ the Lanczos algorithm for open quantum systems to evaluate the quadratic action \eqref{quadratic action in terms of v}. There are multiple approaches to Krylov complexity in open quantum systems. In this work, we employ the generalization of the Lanczos algorithm to the Arnoldi iteration via the Heissenberg form \cite{Bhattacharjee:2022lzy,Bhattacharya:2022gbz,Liu:2022god,Nizami:2023dkf,Nandy:2024evd}, as this framework has proven effective for generating the OTMSS in our previous studies \cite{Li:2024kfm,Li:2024iji,Li:2024ljz,Zhai:2024odw}, particularly building on the methodology established in \cite{Bhattacharya:2022gbz}. We begin by introducing the general Lindblad master equation \cite{Lindblad:1975ef}:
\begin{equation}
	\dot\rho=-i[H,\rho]+\sum_{k}[L_k\rho L_k^\dagger-\frac{1}{2}\{L_k^\dagger L_k,\rho\}],
	\label{general Lindblad}
\end{equation}
where $\rho$ is the density matrix, $H$ is the Hamiltonian, and $L_k^\dagger$ denotes the jump operator (which corresponds to the annihilation operator in our context). Following \cite{Bhattacharjee:2022lzy}, we assume the open system dynamics occurs in an infinite temperature environment-modeled by the maximally mixed state. This serves as a reasonable approximation since the temperature of the inflationary universe is exceedingly high. Consequently, we obtain
\begin{equation}
	\sum_{k}[L_k\rho_\infty L_k^\dagger-\frac{1}{2}\{L_k^\dagger L_k,\rho_\infty\}]=0,
	\label{assumption for rho}
\end{equation} 
where $\rho_\infty$ denotes the infinite temperature-state. Under this assumption, the operator dyamics is governed by 
\begin{equation}
	\mathcal{\hat O}(t)=\exp(i\mathcal{L}t)\mathcal{\hat O},
	\label{dynamic of operator}
\end{equation}
where  the Heisenberg-picture Lindbladian $\mathcal{L}$ is given by \cite{Bhattacharya:2022gbz}, 
\begin{equation}
	\mathcal{L}=\mathcal{L}_H+\mathcal{L}_D,~~\mathcal{L}_H\mathcal{\hat O}=[H,\mathcal{\hat O}],~~\mathcal{L}_D=	\sum_{k}[L_k\mathcal{\hat O} L_k^\dagger-\frac{1}{2}\{L_k^\dagger L_k,\mathcal{\hat O}\}]
\end{equation}
where $\mathcal{L}_H$ governs the Hermitian dynamics, $\mathcal{L}_D$ represents the dissipative part, and $\mathcal{L}$ acts as a superoperator on the operator space. The inner product can be defined as $(A|B) := \mathrm{Tr}[\rho_\infty A^\dagger B]$, which explicitly connects the operator space structure to the Krylov basis.

Secondly, an appropriate basis is required to compute the aforementioned physical quantities. Following \cite{Bhattacharya:2022gbz}, we adopt an orthonormal basis $\{\mathcal{O}_0, \ldots, \mathcal{O}_n, \ldots\}$ generated by the open-system Lindbladian according to the following procedure:
\begin{equation}
	\rm span(\mathcal{O}_0,...,\mathcal{O}_n)=span (\mathcal{V}_0,...,\mathcal{V}_n), 
	\label{span of krylov basis}
\end{equation}
where $\mathcal{V}_n$ is the original Krylov basis, and this basis can be generated by the Arnoldi iteration. Here, we will recall the main idea which is started with an initial normalized vector $\mathcal{O}_0 \propto \mathcal{O}$. Then, we could construct the subsequent Krylov basis vectors by repeatedly applying the operator and orthonormalizing at each step. 
\begin{equation}
	|\mathcal{U}_k)=\mathcal{L}| \mathcal{O}_{k-1}).
\label{recursion relation}
\end{equation}
From $j=0$ to $n-1$, one can perfom the following iterations: $1.~ h_{j,k-1}=(\mathcal{O}_j|\mathcal{U}_k)$, $2.~|\tilde{\mathcal{U}}_k)=|\mathcal{U}_k)-\sum_{j=0}^{k-1}h_{j,k-1}|\mathcal{O}_j)$, $3.~h_{k,k-1}=\sqrt{(\tilde{U}_k | \tilde{\mathcal{U}}_k)}$, it will stop as if $h_{k,k-1}=0$, otherwise, we could define
\begin{equation}
|\mathcal{O}_k)=\frac{|\tilde{\mathcal{U}}_k)}{h_{k,k-1}}.
\label{final recursion}
\end{equation}
Consequently, the Lindbladian can transform into o an upper Heissenberg form in the span Krylov
basis (also called the Arnoldi basis)
\begin{equation}
\mathcal{L}=
\begin{pmatrix}
	h_{0,0} & h_{0,1} & h_{0,2} & \cdots & \cdots & h_{0,n} \\
	h_{1,0} & h_{1,1} & h_{1,2} & \cdots & \cdots & h_{1,n} \\
	0 & h_{2,1} & h_{2,2} & h_{2,3} & \cdots & \cdots \\
	\vdots & 0 & h_{3,2} & \cdots & \cdots & \cdots \\
	0 & \vdots & 0 & \vdots & \ddots & h_{n-1,n} \\
	0 & 0 & \vdots & 0 & h_{n,n-1} & h_{n,n}
\end{pmatrix}
\label{matrix of L}
	\end{equation}
where $h_{m,n}=(\mathcal{O}_m|\mathcal{L}|\mathcal{O}_n)$. When $\mathcal{L}$ is a Hermitian superoperator, the Arnoldi iteration will nicely recover into  the Lanczos algorithm.
Following our previous conventions \cite{Li:2024kfm}, we express the generalized Lanczos algorithm as
\begin{equation}
\mathcal{L}|\mathcal{O}_n)=-ic_n|\mathcal{O}_n)+b_{n+1}|\mathcal{O}_{n+1})+b_n|\mathcal{O}_{n-1}),
\label{eq L in K-basis}
\end{equation}
where the coefficients $c_n$ capture the open quantum system dynamics for a given Hamiltonian (corresponding to the diagonal elements $h_{n,n}$ in \eqref{matrix of L}), while the Lanczos coefficients $b_n$ (associated with the off-diagonal elements $h_{n,n-1}$ in \eqref{matrix of L}) characterize the chaotic behavior of the dynamical system. The orthogonal basis $\mathcal{O}_n = |n,n\rangle_{-\vec{k},\vec{k}}$ spans the Krylov space defined by Eq. \eqref{span of krylov basis}.  The Liouvillian superoperator $\mathcal{L}$, which admits a representation in terms of creation and annihilation operators equivalent to that of a Hamiltonian operator, which governs the time evolution.

This formulation allows us to interpret $\mathcal{L}$ as an effective Hamiltonian that acts on two-mode quantum states, enabling explicit separation between open and closed system contributions. The action of $\mathcal{L}$ on suitable quantum states thus provides a mechanism to:
\begin{enumerate}
	\item Isolate the part of open system
	\item Preserve unitary evolution (closed system)
\end{enumerate}
For comprehensive discussions of the Lanczos algorithm implementation and Krylov complexity in this context, we refer readers to \cite{Parker:2018yvk}.

To connect with the generalized Lanczos algorithm, we need to represent the action \eqref{quadratic action in terms of v} into the Hamiltonian in terms of creation and annihilation operators, where the definition of Hamiltonian is $H=\int d^{3}xd\eta (\pi v'-L)$ with the conjugate momentum as
\begin{equation}
\pi(\eta,\vec{x})=\frac{\delta L}{\delta v'(\eta,\vec{x})}=v'-\frac{a'}{a}v. 
\label{conjugate momentum}
\end{equation}
Therefore, we could obtain the Hamiltonian 
\begin{equation}
H=\frac{1}{2}\int d^{3}xd\eta [\pi^{2}+(\partial_{i}v)^{2}+\frac{a'}{a}(\pi v+v\pi)]. 
\label{Hamiltonian of pi f}
\end{equation}
Utilizing the Fourier decomposition for $v$ and $\pi$, 
\begin{equation}
\hat{v} (\eta,\vec{x})=\int \frac{d^3k}{(2\pi)^{3/2}} \sqrt{\frac{1}{2k}}(\hat{c }_{-\vec{k} }^{\dagger}v_{\vec {k}}^{\ast }e^{-i\vec{k\cdot }\vec{x}}+ {\hat c_{\vec k}}v_{\vec k}e^{i\vec{k\cdot }\vec{x}}),
\label{vk}
\end{equation}
\begin{equation}
\hat{\pi} (\eta,\vec{x})=i\int \frac{d^3k}{(2\pi)^{3/2}}\sqrt{\frac{k}{2}}(\hat{c }_{-\vec{k}}^{\dagger}u_{\vec k}^{\ast }e^{-i\vec{k\cdot }\vec{x}}-\hat{c}_{\vec k}u_{\vec k}e^{i\vec{k\cdot }\vec{x}}).
\label{pik}
\end{equation}
where $\hat{c}_{-\vec{k}}^{\dagger}$ and $\hat{c}_{\vec{k}}$ represent the creation and annihilation operators, respectively, and $u_k(\eta)$, $v_k(\eta)$ are functions of conformal time $\eta$. We retain both modes $(\vec{k}, -\vec{k})$ due to the invariance of the action \eqref{quadratic action in terms of v} under $\vec{k} \rightarrow -\vec{k}$, following the treatment in \cite{Agullo:2022ttg}. Accordingly, the Hamiltonian \eqref{Hamiltonian of pi f} takes the following form:
\begin{equation}
\begin{split}
\hat{H}=\int{d^{3}k}\hat{H}_{k}=\int{d^{3}k}\bigg[k\hat{c }_{-\vec{k} }^{\dagger}\hat{c}_{-\vec{k}}+k\hat{c}_{\vec k}\hat{c}_{\vec k}^{\dagger } +i\frac{z{}'}{z}\hat{c}_{\vec k}^{\dagger }\hat{c}_{-\vec{k} }^{\dagger }-i\frac{z{}'}{z}\hat{c}_{\vec k}\hat{c}_{-\vec{k} }\bigg].
\end{split}
\label{standard hamilton}
\end{equation}
where we have chosen appropriate normalizations for $v_{\vec k}$ and $u_{\vec k}$ and this Hamiltonian is of the $SL(2,R)$ structure \cite{Caputa:2021sib}. We then act with this Hamiltonian \eqref{standard hamilton} on the two-mode orthogonal basis $\mathcal{O}_n=|n,n\rangle_{-\vec{k},\vec{k}}$, yielding: 
\begin{equation}
\begin{split}
\hat H|\mathcal{O}_{n})=(2n+1)k|\mathcal{O}_{n})
+i\frac{z{}'}{z}(n+1)|\mathcal{O}_{n+1})
-in\big(\frac{z{}'}{z}\big)|\mathcal{O}_{n-1}). 
\end{split}
\label{recursive relation in open system}
\end{equation}
where 
\begin{equation}
b_n^2=n^2(\frac{z'}{z})^2,~~~~c_n=i(2n+1)k,
\label{bn and cn}
\end{equation}
The prime symbol denotes differentiation with respect to conformal time. As shown in \cite{Parker:2018yvk}, the dynamical system becomes maximally chaotic and infinite when $b_n\propto n$. Hence, the expression for the Lanczos coefficient in \eqref{bn and cn} clearly demonstrates that the early universe constitutes a maximally chaotic, infinite dynamical system, as established in \cite{Li:2024kfm,Li:2024iji,Li:2024ljz,Zhai:2024odw}. For convenience, we define
\begin{equation}
M_{P}^2|1-\mu_1^2|=\bigg(\frac{z'}{z}\bigg)^2,~~  M_P\mu_2=k,
\label{mu1 and mu2}
\end{equation}
where we follow the notation of \cite{Li:2024kfm} and restore the Planck mass units. Here, $\mu_2$ acts as a dissipative coefficient, leading to the expressions $b_n^2 = n^2|1-\mu_1^2|$ and $c_n = i(2n+1)\mu_2$. As noted earlier, the $c_n$ term corresponds to the open system contribution of the Hamiltonian \eqref{standard hamilton}, while the $b_n$ term aligns with the closed system part. We can therefore decompose the Hamiltonian into open and closed parts as follows:
\begin{equation}
\begin{split}
&H_{open}=M_P\mu_2(\hat{c }_{-k}^{\dagger}\hat{c}_{-k}+\hat{c}_{k}\hat{c}_{k}^{\dagger }),\\&
H_{closed}= iM_P\sqrt{|1-\mu_1^2|}(\hat{c}_{k}^{\dagger }\hat{c}_{-k}^{\dagger }-\hat{c}_{k}\hat{c}_{-k}).
\label{open hamilton}
\end{split}
\end{equation}
For any single field inflationary models, its Hamiltonian can be divided into two parts as shown in Eq. \eqref{open hamilton}, where the information of various models are encoded in $\mu_2$ and $\mu_1$. 

Here, we propose a new perspective on the generalized Lanczos algorithm in this work.
	This algorithm explicitly highlights the open system contribution to the quadratic perturbation action in \eqref{standard hamilton}, an aspect often overlooked in conventional treatments, where the dissipative coefficient is defined in direct relation to various Hamiltonian components.

\section{wave function}
\label{wave function}
As noted in Ref.~\cite{Grishchuk:1990bj}, the amplification from quantum to classical scales during inflation is fundamentally a quantum squeezing process. It is therefore natural to employ the two-mode squeezed state to describe quantum fluctuations in this context. In this work, however, we adopt a more general wave function. Specifically, we use the OTMSS, as introduced in~\cite{Li:2024kfm}, defined as follows:
\begin{widetext}
	\begin{equation}
	|\mathcal{O}(\eta))=e^{i\mathcal{L}\eta}|0,0\rangle_{-\vec{k},\vec k}=\frac{ \rm sech r_k(\eta)}{1+\mu_2\tanh r_k(\eta)}\sum_{n=0}^{\infty}|1-\mu_1^2|^{\frac{n}{2}}\frac{(-\exp(2i\phi_k(\eta))\tanh r_k(\eta))^{n}}{(1+\mu_2\tanh r_k( \eta))^{n}}|n,n\rangle_{-\vec{k},\vec k},
	\label{wave function1}
	\end{equation}
\end{widetext}
where $|n,n\rangle_{-\vec k,\vec k}$ denotes the Arnoldi basis used in this work, $\eta$ is the conformal time, $r_k$ and $\phi_k$ are the squeezing parameters, and $\mu_2$ and $|1-\mu_1^2|$ are as defined in Eq.~\eqref{mu1 and mu2}. We now provide further details on the wave function given in Eq.~\eqref{wave function1}, which describes quantum fluctuations during inflation. This wave function incorporates dissipative effects through the parameter $\mu_2$, and reduces to the standard two-mode squeezed state in the weak dissipative limit ($\mu_2 \ll 1$), 
\begin{widetext}
	\begin{equation}
	|\mathcal{O}(\eta))= \frac{1}{\cosh r_k}\sum_{n=0}^{\infty} (-1)^ne^{2in\phi_k}\tanh^nr_k|n;n\rangle_{\vec{k},-\vec{k}}+\mathcal{O}^n(\mu_2). 
	\label{two mode squeezed state1}
	\end{equation}
\end{widetext}
It follows that the pure two-mode squeezed state corresponds to a wave function describing a closed quantum system, a point we will reaffirm in subsequent discussions. Secondly, the OTMSS given in Eq.~\eqref{wave function1} is constructed based on the second kind of Meixner polynomial, \cite{Hetyei_2009},
\begin{equation}
P_{n+1}(x)=(x-\tilde{c}_n)P_n(x)-b_n^2P_{n-1}(x),
\label{eq definition Pn}
\end{equation}
where we define $\tilde{c}_n=-ic_n$, and $x$ denote the Hamiltonian. The sequence begins with the initial conditions $P_0(x) =1$ and $P_1(x) = x-c_0$. The polynomial $P_n(x)=\det (x-\mathcal{L}_n)$ corresponds to an open system, and $\mathcal{L}_n$ represents the Liouvillian superoperator for the $n$-th quantum state. By introducing the natural orthonormal basis $\{e_n\}$. We could represent $ |P_{n}(x))=\Bigl (\prod_{k=1}^{n}b_{k}\Bigr )|\mathcal{O}_n),\ \ \mbox{and}\ \ \ |x^{n})=\mathcal{L}^{n}|\mathcal{O}).$  Combining Eq. \eqref{eq definition Pn}, there is an explicit relation as follows,
\begin{equation}
b_{n+1}e_{n+1}+b_ne_{n-1}=(x-\tilde{c}_n)e_n,
\label{equalled relation of bn}
\end{equation}
where \( e_n = \mathcal{O}_n \) in our construction, which is equivalent to the generalized Lanczos algorithm given in Eq.~\eqref{eq L in K-basis}. From this straightforward representation, it is evident that the  OTMSS is constructed entirely from the second kind of Meixner polynomial, capturing dissipative effects inherent to open quantum systems such as the early universe. This demonstrates that the OTMSS in Eq.~\eqref{wave function1} is more general than the pure two-mode squeezed state. Moreover, the OTMSS is model-independent, as the parameters \( \mu_2 \) and \( |1 - \mu_1^2| \) encode specific physical information from different cosmological scenarios.

 In this section, we have already checked that the OTMSS is more generic and complete compared to the two-mode squeezed state. Before presenting the conceptual advantage of the OTMSS, we need to provide additional explanation for the two-mode squeezed state. To proceed, we will define a generalized displacement operator as follows:
\begin{equation}
\begin{split}
D(\eta)&\ =e^{iHt}=\exp(ikt(c_{k}c_{k}^{\dagger}+c_{-k}^{\dagger}c_{-k})-\frac{z'}{z}t(c_{-k}^{\dagger}c_{k}^{\dagger}-c_{-k}c_{k}))\\ &\ =\exp(-\eta c_{k}^{\dagger}c_{-k}^{\dagger})\exp(\bar{\eta}c_{-k}c_{k})\exp(-\alpha (c_{k}c_{k}^{\dagger}+c_{-k}^{\dagger}c_{-k})),
\end{split}
\end{equation}
where $t$ is just a parameter, $\hat{H}$ is our Hamiltonian operator \eqref{standard hamilton} and we also define $\eta =\frac{z'}{z}t-i\frac{z'}{z}kt^{2}$ and $\alpha =\frac{1}{2}(\frac{z'}{z}t-\frac{z'}{z}kt^{2})^{2}-ikt$.
After some tedious calculations, we could obtain 
\begin{equation}
\begin{split}
|\mathcal{O}(t))&\ =\exp(-\eta c_{k}^{\dagger}c_{-k}^{\dagger})\exp(\bar{\eta}c_{-k}c_{k})\exp(-\alpha (c_{k}c_{k}^{\dagger}+c_{-k}^{\dagger}c_{-k}))|0,0\rangle_{k,-k}\\
&\ =b\sum_{n=0}^{\infty}\frac{(-1)^{n}}{n!}e^{i2n\phi}\tanh^{n}r(c_{-k}^{\dagger}c_{k}^{\dagger})^{n}|0,0\rangle_{k,-k}\\ &\ =b\sum_{n=0}^{\infty}(-1)^{n}e^{i2n\phi}\tanh^{n}r|n,n\rangle_{k,-k},
\end{split}
\label{calculation of two mode squeezed state}
\end{equation}	
where we have set $\eta =e^{2i\phi}\tanh r$ in complex coordinates, $r=|\eta|$ and $b$ is the normalization factor that can be derived as 
\begin{equation}
1=(\mathcal{O}(t)|\mathcal{O}(t))=|b|^{2}\sum_{n=0}^{\infty}\tanh^{2n}r=|b|^{2}\frac{1}{1-\tanh^{2}r}=|b|^{2}\cosh^{2}r
\label{normalization factor}, 
\end{equation}
we obtain \( b = \frac{1}{\cosh r} \). This result demonstrates that only the \( H_{\text{closed}} \) component constructs the conventional two-mode squeezed state. Consequently, the OTMSS, derived from the second kind of Meixner polynomial, represents a generalization of the standard two-mode squeezed state. To our knowledge, this is the first time the OTMSS has been applied to compute the inflationary power spectrum. This generalization naturally incorporates dissipative effects, as described by the \( H_{\text{open}} \) term in Eq.~\eqref{open hamilton}.

\section{Bogoliubov transformation}
\label{Bogoliubov transformation}
Since the power spectrum of the BD vacuum is in excellent agreement with observational data \cite{Planck:2018jri}, we employ the Bogoliubov transformation to establish a relation between the BD vacuum and the vacuum state defined in Eq.~\eqref{wave function1}. Specifically, the mode function \( v_z(\eta, \vec{k}) \) parameterizes the squeezed vacuum state in a time-dependent cosmological background. The reference mode function associated with the BD vacuum is denoted by \( v_{\text{BD}}(\eta, \vec{k}) \). We then introduce a non-trivial Bogoliubov transformation relating \( v_z(\eta, \vec{k}) \) and \( v_{\text{BD}}(\eta, \vec{k}) \) as follows:
\begin{equation}
v_z(\eta,\vec{k})=\alpha_k v_{BD}(\eta,\vec{k})+\beta_kv_{BD}^*(\eta,\vec{k}),
\label{bogoliubov transformation}
\end{equation}
where the coefficient of Bogoliubov transformation satisfying with the canonical Wronskian condition, 
\begin{equation}
|\alpha_k|^2-|\beta_k|^2=1.
\label{wronskian}
\end{equation}
Meanwhile, the annihilation operator of OTMSS can also be defined by 
\begin{equation}
\hat{c}_{\vec{k}}=\alpha_{k}\hat{a}_{\vec{k}}+\beta_k^*\hat{a}_{-\vec{k}}^{\dagger}.
\label{transformation between c and a}
\end{equation}
Then, we could use the Bogoliubov coefficient to construct the wave function via BD vacuum as follows,
\begin{equation}
|0,0\rangle_{OST} =\frac{1}{\mathcal{N}}\exp\bigg[-\int\frac{d^3 k}{(2\pi)^3}\frac{1}{2}\frac{\beta_k^*}{\alpha_k^*}\hat{a}_k^\dagger\hat{a}_{-k}^\dagger\bigg]|BD\rangle,
\label{relation of vacuums}
\end{equation}
where \(\mathcal{N}\) is the normalization factor, and \(|0,0\rangle_{\text{OST}}\) denotes the vacuum of the OTMSS, which is explicitly constructed from the orthogonal basis \(|n,n\rangle_{-\vec{k},\vec{k}}\) generated via specific recurrence relations \cite{Gamayun:2025hvu,Muck:2022xfc}. The essence of Eq.~\eqref{relation of vacuums} is that the wave function in the target spacetime can be systematically constructed using the Bogoliubov transformation applied to the BD vacuum. It is therefore essential to determine the explicit form of the Bogoliubov coefficients \(\alpha_k\) and \(\beta_k\) in our framework. This is crucial for understanding the correspondence between the squeezed state modes and those of the BD vacuum, as articulated in Eq.~\eqref{bogoliubov transformation}. In practice, we apply the two-mode squeezing operator
	\[
	\mathcal{\hat{S}}_{-k,k} = \exp\left( \frac{r_k}{2} e^{-i\phi_k} \hat{a}_k \hat{a}_{-k} - \frac{r_k}{2} e^{i\phi_k} \hat{a}_k^\dagger \hat{a}_{-k}^\dagger \right)
	\]
	to the annihilation operators of the BD vacuum to derive the Bogoliubov coefficients.
\begin{equation}
\mathcal{\hat S}^\dagger\hat a\mathcal{\hat S}=\hat c_{\vec{k}}=\cosh r_k\hat a_{\vec{k}}-e^{i\phi_k}\sinh r_k\hat a_{-\vec{k}}^\dagger,
\label{bogoliubov transformation}
\end{equation}
Compared with Eq. \eqref{transformation between c and a}, we can obtain the formula for $\alpha_k$ and $\beta_k$
\begin{equation}
\alpha_k=\cosh r_k,~~~\beta_k=-\exp(-i\phi_k)\sinh r_k,
\label{bogoliubov coefficient}
\end{equation}  
where $\phi_k$ and $r_k$ are two parameters depicting the properties for the OTMSS \eqref{wave function1} and the formula of $\alpha_k$ and $\beta_k$ is consistent with \cite{Bianchi:2024jmn,Grain:2019vnq}. Next, we will proceed with the evolution of $\phi_k$ and $r_k$ determined by the Hamiltonian \eqref{standard hamilton}. 

As previously established, the OTMSS in Eq.~\eqref{wave function1} represents a generalization of the pure two-mode squeezed state, as it incorporates dissipative effects inherent to open quantum systems. In this work, we focus specifically on the super-Hubble regime, where $k\eta$ approaches zero. According to \eqref{mu1 and mu2}, $\mu_2$ (the dissipative coefficient) becomes extremely small in this regime, indicating that the universe at the super-Hubble scale behaves as a weakly dissipative system. Therefore, the pure two-mode squeezed state serves as a good approximation, and our derivation for \eqref{bogoliubov coefficient} remains consistent with this understanding. In the sub-horizon regime, as $k\eta$ approaches infinity, the dissipative coefficient $\mu_2$ becomes the unity as discovered by \cite{Zhai:2024odw,Li:2024kfm,Li:2024iji,Li:2024ljz} This indicates that the sub-horizon region acts as a strongly dissipative system, leading to that we should use the OTMSS to investigate the observable by taking place of the two-mode squeezed state. Consequently, we expect observable predictions to deviate from those of the standard scenario. This provides a physical interpretation of observational consequences throughout the entire inflationary epoch: initially, observable outcomes deviate from the standard case in the sub-horizon region. After horizon exit, when the mode enters the super-horizon regime, the observational signatures align with the standard results. In the following calculations, we will use the power spectrum to illustrate this point in the super-horizon regime since it is related to the observations.

 \section{Evolution of $\phi_k$ and $r_k$}
 \label{Evolution of phik and rk}
 To derive the power spectrum of the OTMSS given in Eq.~\eqref{wave function1}, the Bogoliubov coefficients in Eq.~\eqref{bogoliubov coefficient} must be expressed as functions of the wavenumber \(k\), following the approach adopted in Refs.~\cite{Danielsson:2002kx,Danielsson:2004xw,Broy:2016zik}, where certain approximations were employed. To fully incorporate the structure of the OTMSS wave function \eqref{wave function1}, we numerically compute the evolution of the squeezing parameters \(r_k\) and \(\phi_k\) with respect to \(k\). This evolution is governed by the Schr\"{o}dinger equation
 \[
 \hat{H} |\mathcal{O}(\eta)\rangle = i \partial_\eta |\mathcal{O}(\eta)\rangle,
 \]
 which completely determines the dynamics. The resulting equations of motion for \(r_k\) and \(\phi_k\) can be expressed as derived in \cite{Zhai:2024odw}:
 \begin{widetext}
 	\begin{align}
 	&r_k'=\frac{-M_{P}|1-\mu_1^2|\sinh2r_k\cos2\phi_k-\sinh2r_k\cdot \mu_2'}{\sinh2r_k+2\mu_2\cosh^2r_k},\\&
 	\phi_k'=-M_{P}\mu_2+\frac{1}{2}M_{P}|1-\mu_1^2|^{\frac{1}{2}}\sin2\phi_k\Big [|1-\mu_1^2|^{\frac{1}{2}}\frac{\tanh r_k}{1+\mu_2\tanh r_k}+|1-\mu_1^2|^{-\frac{1}{2}}(\coth r_k+\mu_2)\Bigr ].
 	\label{eq rk and phik}
 	\end{align}
 \end{widetext}
Compared to the original evolution equations for \(\phi_k\) and \(r_k\), we have neglected the contribution from the potential term due to the slow-roll conditions. These calculations can be rather involved; for complete details, we refer the reader to the appendix of \cite{Zhai:2024odw}. The evolution equations \eqref{eq rk and phik} for \(r_k\) and \(\phi_k\) describe the squeezing dynamics during the inflationary epoch. The power spectrum is evaluated at horizon crossing for a fiducial scale \(k_* = 0.05~\text{Mpc}^{-1}\).  
To facilitate comparison with observations, we note that CMB measurements cover the range from \(10^{-4}~\text{Mpc}^{-1}\) to \(1~\text{Mpc}^{-1}\) \cite{Planck:2018jri}. It is therefore useful to compute the power spectrum across this full range. Under these considerations, Eq.~\eqref{eq rk and phik} becomes: 
 \begin{equation}
 \begin{split}
 r_k'=&\ -\frac{\tanh r_k(M_{P}|1-\mu_1^2|\cos2\phi_k)}{\tanh r_k+\mu_2},\\
 \phi_k'=&\ -M_{P}\mu_2+\frac{1}{2}\sin2\phi_k\Bigl (\frac{M_{P}|1-\mu_1^2|\tanh r_k}{1+\mu_2\tanh r_k} +M_{P}\coth r_k+M_{P}\mu_2\Bigr ).
 \label{eq of rk and phik 1}
 \end{split}
 \end{equation}
To obtain the corresponding numerical values of \(r_k\) and \(\phi_k\), we use the horizon-crossing condition \(k_* = aH\) at the pivot scale, whose value is inferred from present-day observations. This allows us to express the conformal time as \(\eta = -1/k\) across the CMB observational range \cite{Bianchi:2024qyp,Gong:2001he,Auclair:2022yxs}.  
Using these evolution equations, we numerically compute \(r_k\) and \(\phi_k\) as functions of the comoving momentum \(k\).
 \begin{figure}
 	\centering
 	\includegraphics[width=0.8\linewidth]{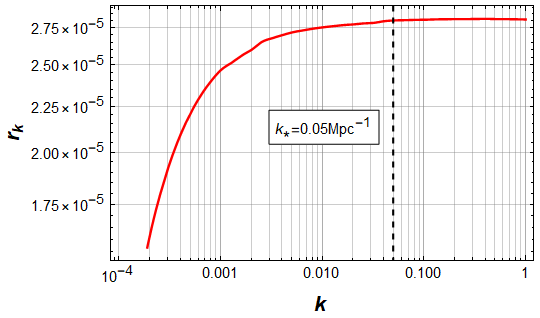}
 	\caption{The numeric of squeezed parameter $r_k$ in terms of $k$ in the horizon crossing. The dotted line is the reference point of $k_*=0.05\ \mbox{Mpc}^{-1}$.}
 	\label{fig:rk1}
 \end{figure}
 \begin{figure}
 	\centering
 	\includegraphics[width=0.75\linewidth]{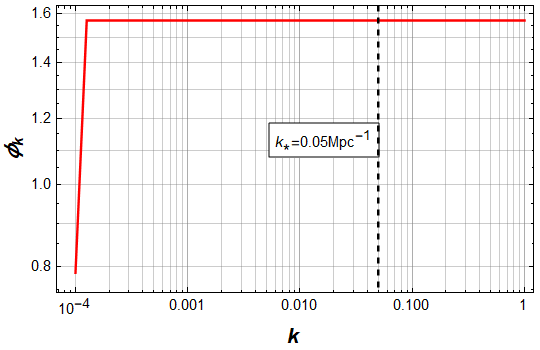}
 	\caption{The numeric of squeezed parameter $\phi_k$ in terms of $k$ in the horizon crossing. The dotted line is the reference point of $k_*=0.05\ \mbox{Mpc}^{-1}$.}
 	\label{fig:phik1}
 \end{figure}
 From Fig. \ref{fig:rk1} and Fig. \ref{fig:phik1}, one can see that the evolution of $r_k$ and $\phi_k$ are almost the constants from $10^{-4}~Mpc^{-1}$ to $1~Mpc^{-1}$. Even, we can further evaluate the number density of OTMSS in terms of  BD vacuum,
 \begin{equation}
 \langle 0,0|\hat c^\dagger_{\vec{k}}\hat{c}_{-\vec{k}}|0,0\rangle= |\beta_k|^2=\sinh^2 r_k,
 \label{number density}
 \end{equation}
 where $|0,0\rangle$ is the BD vacuum and we have implemented Eq. \eqref{transformation between c and a}. In light of this equation, we can give the numeric of mean number density under the BD vacuum. 
 \begin{figure}
 	\centering
 	\includegraphics[width=0.8\linewidth]{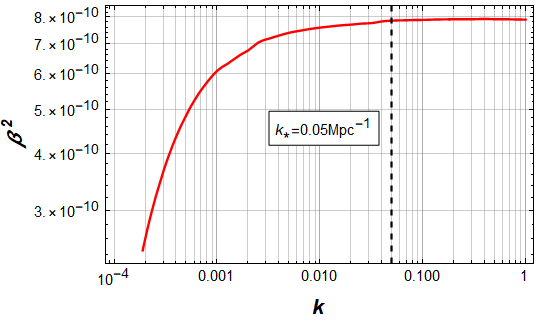}
 	\caption{The numeric of squeezed parameter $|\beta_k|^2$ in terms of $k$ in the horizon crossing. The dotted line is the reference point of $k_*=0.05\ \mbox{Mpc}^{-1}$.}
 	\label{fig:betak}
 \end{figure}
 Fig. \ref{fig:betak} clearly indicates that mean number density is almost zero since the particles are generated in preheating process, although the number density of particle is non-vanishing. Using the evolution of $r_k$ and $\phi_k$ in terms of k, one can investigate the power spectrum.

 \section{Power spectrum of OTMSS}
 \label{Power spectrum of open tw-mode squeezed state}
Since the power spectrum of the OTMSS in Eq.~\eqref{wave function1} scales with \(|v_z|^2\), we may express the ratio between this power spectrum and that of the BD vacuum as follows, using the Bogoliubov transformation given in Eq.~\eqref{bogoliubov transformation}:
 \begin{equation}
 \gamma_z=\frac{\Delta_{\mathcal{R}z}^2}{\Delta_{\mathcal{R}}^2}=|\alpha-\beta|^2=\cosh2r_k+\sinh2r_k\cos\phi_k,
 \label{ratio between BD and sqz}
 \end{equation}
where \(\Delta_{\mathcal{R}z}^2\) denotes the power spectrum for the OTMSS, \(\Delta_{\mathcal{R}}^2\) represents the power spectrum of the BD vacuum, and we have used the super-Hubble scale approximation \(\lim_{k\eta \to 0} v_{BD}^* / v_{BD} = -1\) derived from the BD vacuum definition in Eq.~\eqref{BD vacuum}. With this relation established, it is straightforward to compute the numerical evaluation of Eq.~\eqref{ratio between BD and sqz} as a function of \(k\).
 \begin{figure}
 	\centering
 	\includegraphics[width=0.8\linewidth]{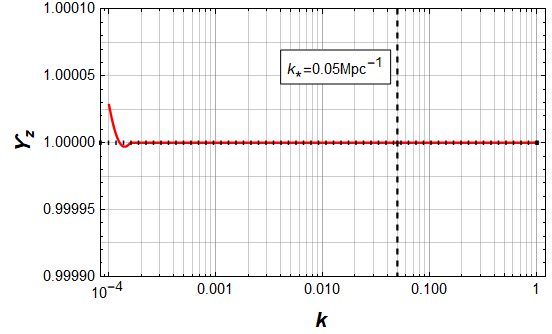}
 	\caption{The numeric of squeezed parameter $\gamma_z(k)$ in terms of $k$ in the horizon crossing. The dotted line is the reference point of $k_*=0.05\ \mbox{Mpc}^{-1}$.}
 	\label{fig:gammak}
 \end{figure}
Fig.~\ref{fig:gammak} shows that the ratio in Eq.~\eqref{ratio between BD and sqz} is very close to unity after horizon crossing, which implies that the power spectrum of the two-mode squeezed state is consistent with that of the BD vacuum. For completeness, we further validate this result using the formal definition of the power spectrum. Following the notation of \cite{Baumann:2009ds}, the two-point correlation function is defined as
 \begin{equation}
 \begin{split}
 \langle|\hat{v}_z|^2\rangle=&\ \langle 0,0|\hat{v}_z^{\dagger}(\eta,0)\hat{v}_z(\eta,0)|0,0\rangle \\
 =&\ \int d\ln{k}\frac{k^3}{2\pi^2}|v_{BD}(\eta,\vec{k})|^2\gamma_z.
 \label{two point function of vz}
 \end{split}
 \end{equation}
 where $\gamma_z$ is the ratio \eqref{ratio between BD and sqz}. Thereafter, we could define the power spectrum of the mode function $v_z$ as follows,
 \begin{equation}
 \begin{split}
 \Delta_{v_z}^2(\eta,k)=&\ \frac{k^3}{2\pi^2}|v_{BD}(\eta,\vec{k})|^2\gamma_z\\
 =&\ \Delta_{v_{BD}}^2(\eta,k)\gamma_z,
 \end{split}
 \label{pwoer spectrum of vz}
 \end{equation}
 To compare with observational data, we compute the power spectrum using the curvature perturbation \(\mathcal{R} = v/z\), where \(z = \sqrt{2\epsilon}\,a\) as defined in Sec.~\ref{Some foundation of inflationary perturbation}. The power spectrum of the curvature perturbation is therefore given by:
 \begin{equation}
 \Delta_{\mathcal{R}z}^2(\eta,k)=\frac{1}{2\epsilon a^2}\frac{\Delta_{v_z}^2(\eta,k)}{M_{P}^2}=\frac{1}{2\epsilon a^2}\frac{\Delta_{v_{BD}}^2\gamma_z(\eta,k)}{M_{P}^2},
 \label{power spectrum of R}
 \end{equation}
As for the power spectrum of the BD vacuum, it takes the form
\[
\Delta_{\mathcal{R}}^2 = A_s \left( \frac{k}{k_*} \right)^{n_s - 1},
\]
where the amplitude \(A_s\) and the spectral index \(n_s\) are constrained by observational data from \cite{Planck:2018jri}.
 \begin{equation}
 A_s=(2.196\pm0.060)\times10^{-9},~~~n_s=0.9649\pm 0.0042.
 \label{observational constraints}
 \end{equation}
 Based on these observational constraints, we could give the prediction for the power spectrum of OTMSS in light of \eqref{power spectrum of R}. 
 \begin{figure}
 	\centering
 	\includegraphics[width=0.8\linewidth]{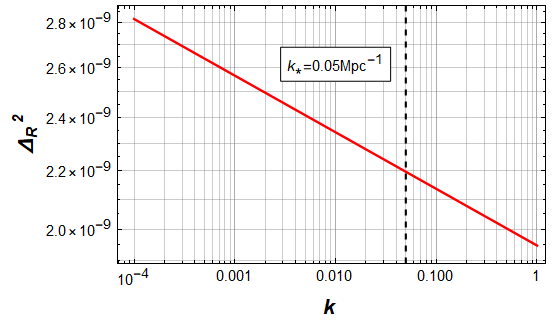}
 	\caption{The numeric of power spectrum $\Delta_{\mathcal{R}z}^2(k)$ in terms of $k$ in the horizon crossing. The dotted line is the reference point of $k_*=0.05\ \mbox{Mpc}^{-1}$.}
 	\label{fig:deltak}
 \end{figure}
Fig.~\ref{fig:deltak} clearly demonstrates that the power spectrum of the OTMSS is nearly identical to that of the BD vacuum, which is consistent with the near-unity ratio shown in Eq.~\eqref{ratio between BD and sqz} and Fig.~\ref{fig:gammak}. Furthermore, our evaluation of the spectral index \(n_s\) agrees with the BD vacuum results within observational uncertainties.

Recalling the calculation of the power spectrum for the squeezed state as outlined in \cite{Danielsson:2002kx,Danielsson:2004xw,Broy:2016zik}, we may parametrize the Bogoliubov coefficients \(\alpha_k\) and \(\beta_k\) in terms of the squeezing parameters \(r_k\) and \(\phi_k\), which are in turn expressed as functions of the momentum \(k\). However, these treatments neglect an essential feature: both \(r_k\) and \(\phi_k\) evolve in time according to the Schrödinger equation. 
A more complete approach must therefore incorporate this dynamical evolution. Moreover, we focus on the OTMSS defined in Eq.~\eqref{wave function1}, which inherently describes an open quantum system---reflecting the fact that the universe is best treated as an open system. Within this framework, our numerical results for the power spectrum align with those of the BD vacuum. Although our predictions are observationally indistinguishable from the BD vacuum, the wave function in Eq.~\eqref{wave function1} is more general: it is model-independent and constructed solely from the second kind of Meixner polynomial within the generalized Lanczos algorithm.

 \section{Conclusion and Discussion}
 \label{conclusion and outlook}
 
Within the framework of single-field inflation, we have developed a method for computing the power spectrum of the two-mode squeezed state. The central elements of this construction are the OTMSS given in Eq.~\eqref{wave function1} and the generalized Lanczos algorithm, as detailed in Sec.~\ref{Lanczos algorithm in inflation}.

Within the framework of the generalized Lanczos algorithm, the Hamiltonian can be decomposed into open and closed system components, as expressed in Eq.~\eqref{open hamilton}. During inflation, an amplification mechanism drives the quantum-to-classical transition, making the two-mode squeezed state a natural basis for characterizing curvature perturbations. This squeezed state is constructed by applying the squeezing operator
\[
\mathcal{\hat{S}}_{-k,k} = \exp\left( u \hat{a}_k \hat{a}_{-k} - \bar{u} \hat{a}_k^\dagger \hat{a}_{-k}^\dagger \right)
\]
to the BD vacuum, where \( u = \frac{r_k}{2} e^{-i\phi_k} \) in our notation. Crucially, Eq.~\eqref{open hamilton} indicates that \( \mathcal{\hat{S}}_{-k,k} \) captures only the closed system dynamics. In contrast, the OTMSS defined in Eq.~\eqref{wave function1} incorporates environmental interactions via the dissipative coefficient \( \mu_2 \). The conventional squeezed state is recovered in the weak dissipation limit (\( \mu_2 \to 0 \)) as the leading-order approximation.

From the Schrödinger equation
\[
\hat{H} |\mathcal{O}(\eta)\rangle = i \partial_\eta |\mathcal{O}(\eta)\rangle,
\]
we derive the evolution equations for \( r_k(\eta) \) and \( \phi_k(\eta) \), with numerical solutions illustrated in Figs.~\ref{fig:phik1} and \ref{fig:rk1}. The ratio \( \gamma_k \) between the OTMSS and BD vacuum power spectra, defined via Bogoliubov transformations in Eq.~\eqref{ratio between BD and sqz}, remains unity throughout inflation, as shown in Fig.~\ref{fig:gammak}. This is consistent with strongly suppressed particle production (mean particle density \( \langle n_k \rangle \ll 1 \)) due to preheating dynamics. Consequently, the power spectrum \( \Delta_{\mathcal{R}}^2(k) \) computed from the OTMSS in Eq.~\eqref{wave function1} matches the prediction of the BD vacuum, as confirmed in Fig.~\ref{fig:deltak}.
 
 \textbf{Key insights:}
 \begin{itemize}
 	\item Dissipation ($\mu_2$) modifies higher-order correlations in the strong dissipative regime but leaves the leading-order power spectrum intact.
 	\item Observational indistinguishability from BD vacuum arises from $\gamma_k \approx 1$, even with $\mu_2 \neq 0$.
 	\item Lanczos-based separation of $\hat{H}$ clarifies the open system's part and closed system's part.
 \end{itemize}

 The derivation of the BD vacuum in Eq.~\eqref{BD vacuum} relies on explicitly solving the mode equation of motion. In contrast, the OTMSS is systematically constructed within the Hamiltonian formalism via the Lanczos algorithm \textit{without solving the equations of motion}; instead, model-dependent information is encoded in the parameters \(\mu_2\) and \(|1 - \mu_1^2|\). Remarkably, the power spectrum predicted by the OTMSS in Eq.~\eqref{wave function1} matches the result from the BD vacuum, indicating that the squeezed-state formulation provides a more fundamental description than direct calculation of the equations of motion.
 
 The key ingredient in this construction is the \textbf{generalized displacement operator} \cite{Caputa:2021sib}, defined as
 \begin{equation}
 \mathcal{D}(\xi) = \exp\left( \xi \hat{H} - \bar{\xi} \hat{H}^\dagger \right),
 \end{equation}  
 where $\hat{H}$ exhibits specific group structures (e.g., the $SL(2.R)$ in our case) and the squeezed operator belongs this kind of operator. When $\hat{H}$ belongs to a Lie group, the operator $\mathcal{D}(\xi)$ provides a rigorous framework for generating quantum states, surpassing traditional EOM-based methods in universality.
 
 \textbf{Extensions and implications:}
 \begin{itemize}
 	\item Current results apply to single-field inflation. Generalizations to multifield scenarios \cite{Liu:2019xhn,Liu:2020zzv,Liu:2020zlr,Liu:2021rgq,Zhang:2022bde} and $f(R)$ gravity \cite{Liu:2018hno,Liu:2018htf} are natural next steps.
 	\item The $\mu_2$-dependence in OTMSS may encode decoherence effects observable in higher-order correlations. Pronounced dissipative effects on sub-horizon scales are expected to cause deviations in non-Gaussian correlators from those predicted by the standard two-mode squeezed state. These significant dissipative effects may also influence decoherence.

 	\item Group-theoretic construction offers a unified approach for quantum fluctuations in curved spacetime.
 \end{itemize}

 \section*{Acknowledgements}
LHL and KHZ are funded by National Natural Science Foundation of China (NSFC) with grant NO. 12165009, Hunan Natural Science Foundation with grant NO. 2023JJ30487 and NO. 2022JJ40340. HQZ is funded by NSFC with grant NO. 12175008.


\section*{References}
\begin{thebibliography}{99}
	


\bibitem{Guth:1980zm}
A.~H.~Guth,
Phys. Rev. D \textbf{23} (1981), 347-356
doi:10.1103/PhysRevD.23.347


\bibitem{Starobinsky:1980te}
A.~A.~Starobinsky,
Phys. Lett. B \textbf{91} (1980), 99-102
doi:10.1016/0370-2693(80)90670-X

\bibitem{Linde:1981mu}
A.~D.~Linde,
Phys. Lett. B \textbf{108} (1982), 389-393
doi:10.1016/0370-2693(82)91219-9


\bibitem{Mukhanov:1990me}
V.~F.~Mukhanov, H.~A.~Feldman and R.~H.~Brandenberger,
Phys. Rept. \textbf{215} (1992), 203-333
doi:10.1016/0370-1573(92)90044-Z


\bibitem{Martin:2024qnn}
J.~Martin, C.~Ringeval and V.~Vennin,
JCAP \textbf{07} (2024), 087
doi:10.1088/1475-7516/2024/07/087
[arXiv:2404.10647 [astro-ph.CO]].


\bibitem{Planck:2018jri}
Y.~Akrami \textit{et al.} [Planck],
Astron. Astrophys. \textbf{641} (2020), A10
doi:10.1051/0004-6361/201833887
[arXiv:1807.06211 [astro-ph.CO]].


\bibitem{CORE:2016ymi}
F.~Finelli \textit{et al.} [CORE],
JCAP \textbf{04} (2018), 016
doi:10.1088/1475-7516/2018/04/016
[arXiv:1612.08270 [astro-ph.CO]].


\bibitem{CMB-S4:2020lpa}
K.~Abazajian \textit{et al.} [CMB-S4],
Astrophys. J. \textbf{926} (2022) no.1, 54
doi:10.3847/1538-4357/ac1596
[arXiv:2008.12619 [astro-ph.CO]].

\bibitem{LiteBIRD:2022cnt}
E.~Allys \textit{et al.} [LiteBIRD],
PTEP \textbf{2023} (2023) no.4, 042F01
doi:10.1093/ptep/ptac150
[arXiv:2202.02773 [astro-ph.IM]].


\bibitem{Euclid:2023shr}
M.~Ballardini \textit{et al.} [Euclid],
Astron. Astrophys. \textbf{683} (2024), A220
doi:10.1051/0004-6361/202348162
[arXiv:2309.17287 [astro-ph.CO]].


\bibitem{Mergulhao:2023ukp}
T.~Mergulh\~ao, F.~Beutler and J.~A.~Peacock,
JCAP \textbf{08} (2023), 012
doi:10.1088/1475-7516/2023/08/012
[arXiv:2303.13946 [astro-ph.CO]].


\bibitem{Antony:2024vrx}
A.~Antony, F.~Finelli, D.~K.~Hazra, D.~Paoletti and A.~Shafieloo,
[arXiv:2403.19575 [astro-ph.CO]].

\bibitem{Balitsky:2014epa}
J.~V.~Balitsky and V.~V.~Kiselev,
Phys. Rev. D \textbf{90} (2014) no.12, 125017
doi:10.1103/PhysRevD.90.125017
[arXiv:1406.3046 [gr-qc]].

\bibitem{Dalianis:2024}
I.~Dalianis, ``Features in the Inflaton Potential and the Spectrum of Cosmological Perturbations,'' In: Papantonopoulos, E., Mavromatos, N. (eds) Compact Objects in the Universe. Springer, Cham. https://doi.org/10.1007/978-3-031-55098-0\_13

\bibitem{Papadimitriou:2003}
C.~H.~Papadimitriou, ``Computational complexity." Encyclopedia of computer science. 2003. 260-265.

\bibitem{Parker:2018yvk}
D.~E.~Parker, X.~Cao, A.~Avdoshkin, T.~Scaffidi and E.~Altman,
Phys. Rev. X \textbf{9} (2019) no.4, 041017
doi:10.1103/PhysRevX.9.041017
[arXiv:1812.08657 [cond-mat.stat-mech]].

\bibitem{Choudhury:2020hil}
S.~Choudhury, S.~Chowdhury, N.~Gupta, A.~Mishara, S.~P.~Selvam, S.~Panda, G.~D.~Pasquino, C.~Singha and A.~Swain,
Symmetry \textbf{13} (2021) no.7, 1301
doi:10.3390/sym13071301
[arXiv:2012.10234 [hep-th]].



\bibitem{Bhargava:2020fhl}
P.~Bhargava, S.~Choudhury, S.~Chowdhury, A.~Mishara, S.~P.~Selvam, S.~Panda and G.~D.~Pasquino,
SciPost Phys. Core \textbf{4} (2021), 026
doi:10.21468/SciPostPhysCore.4.4.026
[arXiv:2009.03893 [hep-th]].



\bibitem{Bhattacharyya:2020rpy}
A.~Bhattacharyya, S.~Das, S.~Shajidul Haque and B.~Underwood,
Phys. Rev. D \textbf{101} (2020) no.10, 106020
doi:10.1103/PhysRevD.101.106020
[arXiv:2001.08664 [hep-th]].


\bibitem{Adhikari:2021ked}
K.~Adhikari, S.~Choudhury, H.~N.~Pandya and R.~Srivastava,
Symmetry \textbf{15} (2023) no.3, 664
doi:10.3390/sym15030664
[arXiv:2108.10334 [gr-qc]].

\bibitem{Li:2023ekd}
T.~Li and L.~H.~Liu,
Phys. Lett. B \textbf{854} (2024), 138728
doi:10.1016/j.physletb.2024.138728
[arXiv:2309.01595 [gr-qc]].


\bibitem{Liu:2021nzx}
L.~H.~Liu and A.~C.~Li,
Phys. Dark Univ. \textbf{37} (2022), 101123
doi:10.1016/j.dark.2022.101123
[arXiv:2102.12014 [gr-qc]].

\bibitem{Li:2021kfq}
A.~c.~Li, X.~F.~Li, D.~f.~Zeng and L.~H.~Liu,
Phys. Dark Univ. \textbf{43} (2024), 101422
doi:10.1016/j.dark.2024.101422
[arXiv:2102.12939 [gr-qc]].


\bibitem{Adhikari:2022oxr}
K.~Adhikari and S.~Choudhury,
Fortsch. Phys. \textbf{70} (2022) no.12, 2200126
doi:10.1002/prop.202200126
[arXiv:2203.14330 [hep-th]].


\bibitem{Li:2024kfm}
T.~Li and L.~H.~Liu,
JHEP \textbf{04} (2024), 123
doi:10.1007/JHEP04(2024)123
[arXiv:2401.09307 [hep-th]].


\bibitem{Zhai:2024odw}
K.~H.~Zhai and L.~H.~Liu,
[arXiv:2411.18405 [hep-th]].


\bibitem{Kowalczyk:2024ech}
M.~Kowalczyk and G.~A.~Mena Marug\'an,
doi:10.1103/PhysRevD.110.103502
[arXiv:2409.15886 [gr-qc]].


\bibitem{Vilenkin:1982wt}
A.~Vilenkin and L.~H.~Ford,
Phys. Rev. D \textbf{26} (1982), 1231
doi:10.1103/PhysRevD.26.1231


\bibitem{Starobinsky:2001kn}
A.~A.~Starobinsky,
Pisma Zh. Eksp. Teor. Fiz. \textbf{73} (2001), 415-418
doi:10.1134/1.1381588
[arXiv:astro-ph/0104043 [astro-ph]].

\bibitem{Gessey-Jones:2021yky}
T.~Gessey-Jones and W.~J.~Handley,
Phys. Rev. D \textbf{104} (2021) no.6, 063532
doi:10.1103/PhysRevD.104.063532
[arXiv:2104.03016 [astro-ph.CO]].

\bibitem{Handley:2016ods}
W.~J.~Handley, A.~N.~Lasenby and M.~P.~Hobson,
Phys. Rev. D \textbf{94} (2016) no.2, 024041
doi:10.1103/PhysRevD.94.024041
[arXiv:1607.04148 [gr-qc]].

\bibitem{Contaldi:2003zv}
C.~R.~Contaldi, M.~Peloso, L.~Kofman and A.~D.~Linde,
JCAP \textbf{07} (2003), 002
doi:10.1088/1475-7516/2003/07/002
[arXiv:astro-ph/0303636 [astro-ph]].


\bibitem{Bunch:1978yq}
T.~S.~Bunch and P.~C.~W.~Davies,
Proc. Roy. Soc. Lond. A \textbf{360} (1978), 117-134
doi:10.1098/rspa.1978.0060

\bibitem{Baumann:2009ds}
D.~Baumann,
doi:10.1142/9789814327183\_0010
[arXiv:0907.5424 [hep-th]].



\bibitem{Unruh:1976db}
W.~G.~Unruh,
Phys. Rev. D \textbf{14} (1976), 870
doi:10.1103/PhysRevD.14.870


\bibitem{Agullo:2022ttg}
I.~Agullo, B.~Bonga and P.~R.~Metidieri,
JCAP \textbf{09} (2022), 032
doi:10.1088/1475-7516/2022/09/032
[arXiv:2203.07066 [gr-qc]].


\bibitem{Chernikov:1968zm}
N.~A.~Chernikov and E.~A.~Tagirov,
Ann. Inst. H. Poincare A Phys. Theor. \textbf{9} (1968), 109

\bibitem{Tagirov:1972vv}
E.~A.~Tagirov,
Annals Phys. \textbf{76} (1973), 561-579
doi:10.1016/0003-4916(73)90047-X







\bibitem{Li:2024iji}
T.~Li and L.~H.~Liu,
[arXiv:2405.01433 [hep-th]].


\bibitem{Li:2024ljz}
T.~Li and L.~H.~Liu,
[arXiv:2408.03293 [hep-th]].



\bibitem{Shandera:2017qkg}
S.~Shandera, N.~Agarwal and A.~Kamal,
Phys. Rev. D \textbf{98} (2018) no.8, 083535
doi:10.1103/PhysRevD.98.083535
[arXiv:1708.00493 [hep-th]].


\bibitem{Cheung:2007st}
C.~Cheung, P.~Creminelli, A.~L.~Fitzpatrick, J.~Kaplan and L.~Senatore,
JHEP \textbf{03} (2008), 014
doi:10.1088/1126-6708/2008/03/014
[arXiv:0709.0293 [hep-th]].


\bibitem{Kofman:1997yn}
L.~Kofman, A.~D.~Linde and A.~A.~Starobinsky,
Phys. Rev. D \textbf{56} (1997), 3258-3295
doi:10.1103/PhysRevD.56.3258
[arXiv:hep-ph/9704452 [hep-ph]].


\bibitem{Kofman:1994rk}
L.~Kofman, A.~D.~Linde and A.~A.~Starobinsky,
Phys. Rev. Lett. \textbf{73} (1994), 3195-3198
doi:10.1103/PhysRevLett.73.3195
[arXiv:hep-th/9405187 [hep-th]].


\bibitem{Bhattacharjee:2022lzy}
B.~Bhattacharjee, X.~Cao, P.~Nandy and T.~Pathak,
JHEP \textbf{03} (2023), 054
doi:10.1007/JHEP03(2023)054
[arXiv:2212.06180 [quant-ph]].



\bibitem{Bhattacharya:2022gbz}
A.~Bhattacharya, P.~Nandy, P.~P.~Nath and H.~Sahu,
JHEP \textbf{12} (2022), 081
doi:10.1007/JHEP12(2022)081
[arXiv:2207.05347 [quant-ph]].


\bibitem{Liu:2022god}
C.~Liu, H.~Tang and H.~Zhai,
Phys. Rev. Res. \textbf{5} (2023) no.3, 033085
doi:10.1103/PhysRevResearch.5.033085
[arXiv:2207.13603 [cond-mat.str-el]].

\bibitem{Nizami:2023dkf}
A.~A.~Nizami and A.~W.~Shrestha,
Phys. Rev. E \textbf{108} (2023) no.5, 5
doi:10.1103/PhysRevE.108.054222
[arXiv:2305.00256 [quant-ph]].


\bibitem{Nandy:2024evd}
P.~Nandy, A.~S.~Matsoukas-Roubeas, P.~Mart{\'\i}nez-Azcona, A.~Dymarsky and A.~del Campo,
Phys. Rept. \textbf{1125-1128} (2025), 1-82
doi:10.1016/j.physrep.2025.05.001
[arXiv:2405.09628 [quant-ph]].

\bibitem{Lindblad:1975ef}
G.~Lindblad,
Commun. Math. Phys. \textbf{48} (1976), 119
doi:10.1007/BF01608499





\bibitem{Parker:2018yvk}
D.~E.~Parker, X.~Cao, A.~Avdoshkin, T.~Scaffidi and E.~Altman,
Phys. Rev. X \textbf{9} (2019) no.4, 041017
doi:10.1103/PhysRevX.9.041017
[arXiv:1812.08657 [cond-mat.stat-mech]].


\bibitem{Grishchuk:1990bj}
L.~P.~Grishchuk and Y.~V.~Sidorov,
Phys. Rev. D \textbf{42} (1990), 3413-3421
doi:10.1103/PhysRevD.42.3413

\bibitem{Hetyei_2009}
H. G{\'a}bor, 
Proceedings of the Royal Society A: Mathematical, Physical and Engineering Sciences, \textbf{10} (2009), no.2117, 1471-2946,
doi:10.1098/rspa.2009.0497
[arXiv:0909.4352[Mathematics-Quantum Algebra]].



\bibitem{Gamayun:2025hvu}
O.~Gamayun, M.~A.~Mir, O.~Lychkovskiy and Z.~Ristivojevic,
JHEP \textbf{07} (2025), 256
doi:10.1007/JHEP07(2025)256
[arXiv:2504.03435 [quant-ph]].


\bibitem{Muck:2022xfc}
W.~M{\"u}ck and Y.~Yang,
Nucl. Phys. B \textbf{984} (2022), 115948
doi:10.1016/j.nuclphysb.2022.115948
[arXiv:2205.12815 [hep-th]].



\bibitem{Bianchi:2024jmn}
E.~Bianchi and M.~Gamonal,
[arXiv:2410.11812 [gr-qc]].


\bibitem{Grain:2019vnq}
J.~Grain and V.~Vennin,
JCAP \textbf{02} (2020), 022
doi:10.1088/1475-7516/2020/02/022
[arXiv:1910.01916 [astro-ph.CO]].



\bibitem{Danielsson:2002kx}
U.~H.~Danielsson,
Phys. Rev. D \textbf{66} (2002), 023511
doi:10.1103/PhysRevD.66.023511
[arXiv:hep-th/0203198 [hep-th]].


\bibitem{Danielsson:2004xw}
U.~H.~Danielsson,
Phys. Rev. D \textbf{71} (2005), 023516
doi:10.1103/PhysRevD.71.023516
[arXiv:hep-th/0411172 [hep-th]].

\bibitem{Broy:2016zik}
B.~J.~Broy,
Phys. Rev. D \textbf{94} (2016) no.10, 103508
doi:10.1103/PhysRevD.94.103508
[arXiv:1609.03570 [hep-th]].


\bibitem{Bianchi:2024qyp}
E.~Bianchi and M.~Gamonal,
Phys. Rev. D \textbf{110} (2024) no.10, 104032
doi:10.1103/PhysRevD.110.104032
[arXiv:2405.03157 [gr-qc]].



\bibitem{Gong:2001he}
J.~O.~Gong and E.~D.~Stewart,
Phys. Lett. B \textbf{510} (2001), 1-9
doi:10.1016/S0370-2693(01)00616-5
[arXiv:astro-ph/0101225 [astro-ph]].


\bibitem{Auclair:2022yxs}
P.~Auclair and C.~Ringeval,
Phys. Rev. D \textbf{106} (2022) no.6, 063512
doi:10.1103/PhysRevD.106.063512
[arXiv:2205.12608 [astro-ph.CO]].

\bibitem{Caputa:2021sib}
P.~Caputa, J.~M.~Magan and D.~Patramanis,
Phys. Rev. Res. \textbf{4} (2022) no.1, 013041
doi:10.1103/PhysRevResearch.4.013041
[arXiv:2109.03824 [hep-th]].


\bibitem{Liu:2019xhn}
L.~H.~Liu and W.~L.~Xu,
Chin. Phys. C \textbf{44} (2020) no.8, 085103
doi:10.1088/1674-1137/44/8/085103
[arXiv:1911.10542 [astro-ph.CO]].

\bibitem{Liu:2020zzv}
L.~H.~Liu and T.~Prokopec,
JCAP \textbf{06} (2021), 033
doi:10.1088/1475-7516/2021/06/033
[arXiv:2005.11069 [astro-ph.CO]].

\bibitem{Liu:2020zlr}
L.~H.~Liu, B.~Liang, Y.~C.~Zhou, X.~D.~Liu, W.~L.~Xu and A.~C.~Li,
Phys. Rev. D \textbf{103} (2021) no.6, 063515
doi:10.1103/PhysRevD.103.063515
[arXiv:2007.08278 [astro-ph.CO]].

\bibitem{Liu:2021rgq}
L.~H.~Liu,
Chin. Phys. C \textbf{47} (2023) no.1, 015105
doi:10.1088/1674-1137/ac9d28
[arXiv:2107.07310 [astro-ph.CO]].

\bibitem{Zhang:2022bde}
X.~z.~Zhang, L.~h.~Liu and T.~Qiu,
Phys. Rev. D \textbf{107} (2023) no.4, 043510
doi:10.1103/PhysRevD.107.043510
[arXiv:2207.07873 [hep-th]].


\bibitem{Liu:2018hno}
L.~H.~Liu, T.~Prokopec and A.~A.~Starobinsky,
Phys. Rev. D \textbf{98} (2018) no.4, 043505
doi:10.1103/PhysRevD.98.043505
[arXiv:1806.05407 [gr-qc]].


\bibitem{Liu:2018htf}
L.~H.~Liu,
doi:10.1007/s10773-018-3809-0
[arXiv:1807.00666 [gr-qc]].














	
	




\end{thebibliography}
\end{document}